\documentclass{article}
\usepackage{amsfonts,amssymb, latexsym, amscd}
\title{Tau-functions and special solutions  
in a coupled Painlev\'{e} 
system\thanks{Isaac Newton Institute preprint NI02001-ITS.} 
}
\author{A.N.W. Hone  \\
\normalsize
Institute of Mathematics \& Statistics, \\
\normalsize
 University of Kent at Canterbury, \\
\normalsize
 Kent, U.K.}  

\begin{document}
\def\underset#1#2{\mathrel{\mathop{#2}\limits_{#1}}}
\renewcommand{\theequation}{\arabic{section}.\arabic{equation}}
\newcommand{\beq}{\begin{equation}}
\newcommand{\eeq}{\end{equation}}
\newcommand{\bea}{\begin{eqnarray}}
\newcommand{\eea}{\end{eqnarray}}
\newcommand{\ba}{\begin{array}}
\newcommand{\ea}{\end{array}}
\maketitle

\begin{abstract}
For a pair of coupled Painlev\'e equations obtained as a scaling 
similarity reduction of the Hirota-Satsuma system we describe 
special parameter-families of solutions given in terms of 
mixtures of rational and Airy functions, and in terms of a  
second Painlev\'e transcendent. The tau-functions associated 
to some of these solutions are also given explicitly.   
\end{abstract}

\section{Introduction} 

In a recent pair of articles \cite{hone1,hone2} we have considered 
the following pair of coupled Painlev\'e equations, 
\beq
\ba{ccc}
L_1L_1''-\frac{1}{2}(L_1')^2+(L_1+3L_2+2z)L_1^2+\frac{1}{2}\ell_1^2
& = & 0, \\ && \\
L_2L_2''-\frac{1}{2}(L_2')^2+(3L_1+L_2+2z)L_2^2+\frac{1}{2}\ell_2^2 & = 
& 0,  

\ea
\label{eq:cup34}
\eeq
which arise as a scaling similarity reduction of the Hirota-Satsuma 
system of partial differential equations.  
The system (\ref{eq:cup34}) 
is a coupling between two copies of the equation
P34 in Ince's classification \cite{ince}, to which it clearly degenerates
via the consistent
reductions
$L_1=0=\ell_1$ and $L_2=0=\ell_2$. 
The similarity reduction was 
originally found in our thesis \cite{thesis}, 
but our interest in the pair of equations (\ref{eq:cup34})  
was further stimulated by a conjecture that it  
is connected to a fifth order equation appearing in a classification 
of higher order Painlev\'e equations made by Cosgrove \cite{cos}.  

The Hirota-Satsuma system is 
itself a 4-reduction of the KP hierarchy \cite{hir}, 
which means that the pair of equations (\ref{eq:cup34}) can be derived 
from an $sl(4)$ isomonodromic Lax pair. In our first article \cite{hone1} 
we presented B\"acklund transformations (BTs) 
for the system (\ref{eq:cup34}), 
which could be interpreted in terms of a 
subgroup of the affine Weyl group $W(A_3)$ acting 
on the space of parameters $(\ell_1,\ell_2)$. In that article we 
also described families of special lines in the $(\ell_1,\ell_2)$ 
plane where the system admits two-parameter and three-parameter 
families of solutions.   

The description of the BTs for the system (\ref{eq:cup34}) is much 
simplified by rewriting it as four coupled first order equations, 
namely 
\beq
\ba{ccl}
X_1' & = & -\frac{1}{2}L_1-\frac{3}{2}L_2-X_1^2-z, \\
&& \\
X_2' & = & -\frac{3}{2}L_1-\frac{1}{2}L_2-X_2^2-z, \\
&& \\
L_1' & = & 2L_1X_1-\ell_1, \\
&& \\
L_2' & = & 2L_2X_2-\ell_2.
\ea
\label{eq:sys3}
\eeq
Below we will present these BTs once again, and then show how they 
can be applied to obtain special families of solutions at isolated 
points in the parameter space. The second work \cite{hone2} 
was primarily concerned with the tau-functions for the system 
(\ref{eq:cup34}) (or equivalently (\ref{eq:sys3})), 
and the multilinear lattice equations connecting them. In what 
follows we give exact expressions for the tau-functions of some 
of the special solutions. 

\section{B\"acklund transformations} 

\setcounter{equation}{0}

BTs for the coupled Painlev\'e equations 
(\ref{eq:cup34}) were first found in 
\cite{hone1}, but were presented more       
explicitly in terms of the variables         
of the system (\ref{eq:sys3}) in \cite{hone2}. There are 
two basic reflections in the $(\ell_1,\ell_2)$ plane: 
$$
R: \quad (\ell_1,\ell_2)\to(\ell_2,\ell_1), \qquad
X_1\leftrightarrow X_2, \qquad L_1\leftrightarrow L_2.
$$
$$
S: \quad (\ell_1,\ell_2)\to(-\ell_1,\ell_2),
\quad X_1 \to X_1^{\dagger}:=X_1-\frac{\ell_1}{L_1}, \quad
 X_2 \to  X_2, \quad L_1 \to L_1, \quad L_2 \to L_2.
$$
By combining the two reflectional symmetries $R,S$ it is straightforward
to obtain BTs connecting solutions of the system (\ref{eq:sys3})
at all the points
\beq
(\epsilon \ell_1,\epsilon '\ell_2),\qquad
(\epsilon \ell_2,\epsilon '\ell_1),\qquad \epsilon,\epsilon '=\pm 1
\label{eq:pts}
\eeq
in the parameter space. 

The affine symmetry of the solutions is generated by a   
translational BT, denoted $T$. 
For convenience
we introduce the vector notation
$$
{\bf l}=\left( \begin{array}{cc} \ell_1 \\ \ell_2 \end{array}
\right),
\qquad
{\bf c}=\left( \begin{array}{cc} 2 \\ 2 \end{array}
\right).
$$ 
Then $T$ is defined thus: 
$$
T: \qquad \qquad {\bf l}\to{\bf l}+{\bf c},
\quad
X_1\to\overline{X}_1:=X_2^{\dagger}
-\frac{(\ell_1+\ell_2+2)}{(L_1+L_2+2X_1^{\dagger}X_2^{\dagger}+2z)},
$$
\beq
X_2\to\overline{X}_2:=X_1^{\dagger}
-\frac{(\ell_1+\ell_2+2)}{(L_1+L_2+2X_1^{\dagger}X_2^{\dagger}+2z)},
\label{eq:tra}
\eeq
$$
L_1\to\overline{L}_1:=L_2+\frac{\ell_1+\ell_2+2}
{L_1+L_2+2X_1^{\dagger}X_2^{\dagger}+2z}\left(
2X_1^\dagger-
\frac{(\ell_1+\ell_2+2)}{(L_1+L_2+2X_1^{\dagger}X_2^{\dagger}+2z)}
\right),
$$
$$
L_2\to\overline{L}_2:=L_1+\frac{\ell_1+\ell_2+2}
{L_1+L_2+2X_1^{\dagger}X_2^{\dagger}+2z}\left(
2X_2^\dagger-
\frac{(\ell_1+\ell_2+2)}{(L_1+L_2+2X_1^{\dagger}X_2^{\dagger}+2z)}
\right).
$$
This transformation is explicitly invertible;  
the exact expression for 
$T^{-1}$ may be found in \cite{hone2}.   

\section{Hamiltonians and tau-functions} 

\setcounter{equation}{0}

Part of our original motivation 
for deriving the system (\ref{eq:cup34}) was that it 
may be written as a non-autonomous version of 
an integrable Hamiltonian system of two particles 
interacting via a quartic potential. More precisely, 
(\ref{eq:cup34}) arises from the Hamiltonian 
\beq
h=\frac{1}{2}(p_1^2+p_2^2)+\frac{1}{8}(q_1^4+6q_1^2q_2^2+q_2^4)
+\frac{1}{2}z(q_1^2+q_2^2)
-\frac{1}{8}\left(\frac{\ell_1^2}{q_1^2}+ \frac{\ell_2^2}{q_2^2}\right)
\label{eq:34ham}
\eeq
by making the transformation to canonical conjugate coordinates
and momenta according to
$$
L_j=q_j^2, \qquad L_j'=2p_jq_j, \qquad j=1,2.
$$
It turns out that this system is related by a canonical (contact) 
transformation to another Hamiltonian with quartic potential, 
namely 
\beq
H=\frac{1}{2}(P_1^2+P_2^2)-\frac{1}{16}(Q_1^4+6Q_1^2Q_2^2+8Q_2^4)
+\frac{1}{2}z(Q_1^2+4Q_2^2)-\frac{2\nu^2}{Q_1^2}-\xi Q_2,  
\label{eq:234ham}
\eeq
with the parameters related by 
$$ 
{\bf m[l]}:=\left( \begin{array}{c} \nu \\ \xi \end{array} \right) 
=\left( \begin{array}{c} \ell_1+\ell_2-2 \\
\ell_1-\ell_2 \end{array} \right). 
$$ 
The autonomous versions of the Hamiltonians 
(\ref{eq:34ham},\ref{eq:234ham}) were considered in \cite{fbe}. 

Hamilton's equations for (\ref{eq:234ham}) lead to another 
coupled Painlev\'e system, which is a coupling between P34 and the 
second Painlev\'e equation P2. Indeed, simple reflectional 
symmetries of this other system lead to a derivation 
of the translational BT $T$ for the original 
system (\ref{eq:cup34}) (see \cite{hone1,hone2} for 
further details). Here we simply wish to note that the Hamiltonians 
(\ref{eq:34ham},\ref{eq:234ham}) can be expressed as 
logarithmic derivatives of holomorphic tau-functions for the system.  
This is the analogue in this higher order setting 
of Okamoto's results \cite{hamok,oka} 
on the tau-functions and Hamiltonian structures of the 
Painlev\'e equations. Associated to the first Hamiltonian 
(\ref{eq:34ham}) is a tau-function, denoted $\tau$ in \cite{hone2}, 
such that 
\beq
h_{\bf l}(z)=\frac{d}{dz}\log \tau_{\bf l}(z)   
\label{eq:tau1}
\eeq
(the subscripts signify the dependence on the parameters). 
Corresponding to the second Hamiltonian (\ref{eq:234ham}) related 
by a canonical transformation there is another 
tau-function, denoted $\rho$, with 
\beq 
H_{\bf m[l]}(z)=-2\frac{d}{dz}\log \rho_{\bf m[l]} (z).  
\label{eq:tau2}
\eeq
In terms of the variables in the coupled system (\ref{eq:sys3}), 
these two Hamiltonians are related by the formula  
$$
H_{\bf m[l]}=-4h_{\bf l}-2(X_1+X_2).
$$
 
By making use of the BTs we derived bilinear and multilinear lattice 
equations for the tau-functions at different points in parameter 
space, and expressed the variables $L_j,X_j$ explicitly in terms 
of them. The reader is referred to \cite{hone2} for the exact 
formulae. 

\section{Special lines and special points} 

We expect that the solutions of the system (\ref{eq:cup34}) 
define new transcendental functions at generic points in 
parameter space, and this is expectation is 
supported by work of Cosgrove \cite{cos}. However, in \cite{hone1} 
we showed that along three  
families of lines (denoted $\mathcal{L}_j$) 
in the $(\ell_1,\ell_2)$ plane the system admits 
parameter families of 
solutions in terms of known (P2, or equivalently P34) transcendents.       

\begin{itemize} 
\item $\mathcal{L}_1$: Along the lines
$$
\ell_1=2n \qquad and \qquad \ell_2=2n, \qquad n\in {\mathbb Z},
$$
there is a three-parameter family of special solutions to
the system (\ref{eq:sys3}), obtained by applying the BTs $R,S$ and $T$ 
to the reduction $L_2=0$ on the line $\ell_2=0$, with $L_1$ satisfying 
P34. 

\item $\mathcal{L}_2$: On the lines
$$
\ell_1 \pm \ell_2=4n, \qquad n\in {\mathbb Z},
$$
there is a two-parameter family of special solutions, obtained by
application of the BTs to the reduction $L_1=L_2=L$ on the line 
$\ell_1-\ell_2=0$, where $L$ satisfies P34 with a different scaling 
to the $\mathcal{L}_1$ case above. 

\item $\mathcal{L}_3$: On the lines
$$
\ell_1 \pm \ell_2=2(2n+1), \qquad n\in {\mathbb Z},
$$
there is a three-parameter family of special solutions, generated by
application of the BTs to a special reduction on the line 
$\ell_1+\ell_2=2$, which corresponds to the reduction 
$Q_1=0=\nu$ in the second                
Hamiltonian system (\ref{eq:234ham}), with $Q_2$ satisfying P2. 
\end{itemize} 

In \cite{hone2} we described how 
particular rational solutions could be found 
at the points 
$$ 
(\ell_1,\ell_2)=(2(m+n)+\epsilon/2, 2(m-n)+\epsilon'/2), \qquad
\epsilon,\epsilon'=\pm 1, \qquad
(m,n)\in {\mathbb Z}^2
$$ 
on the lines $\mathcal{L}_2$. In \cite{hone1} we mentioned that 
solutions in terms of mixtures of rational and Airy 
functions could be obtained at 
the intersections of the families  $\mathcal{L}_j$. Below we consider 
these intersections more explicitly. 

\subsection{Intersection of  $\mathcal{L}_1$ and  $\mathcal{L}_2$} 

A result originally appearing in our thesis \cite{thesis} was 
that at the point $(0,0)$ in parameter space a separation of 
variables is possible and the general (four-parameter) 
solution of the system (\ref{eq:cup34}) is given in terms of two copies 
of P2 with zero parameter. More precisely, the variables in the 
system (\ref{eq:sys3}) are given by 
$$ 
L_1=-(y_++y_-)^2, \quad 
L_2=-(y_+-y_-)^2, \quad 
X_1=\left(\log (y_++y_-)\right)', \quad 
X_2=\left(\log (y_+-y_-)\right)',             
$$ 
where $y_\pm$ are two solutions of  
\beq
y''-2y^3+zy=0. 
\label{eq:p2} 
\eeq 
In terms of P2 tau-functions $T, \overline{T}$ related by a 
B\"acklund transformation, the solution of (\ref{eq:p2}) is expressed 
as 
$$ 
y=\left(\log \overline{T}/T \right)'. 
$$ 
Then it turns out that the tau-function $\tau$ for the system 
(\ref{eq:sys3}) at point ${\bf 0}=(0,0)^T$ in parameter space is given by a 
product of four P2 tau-functions, 
$$ 
\tau_{\bf 0}=T_+T_-\overline{T}_+\overline{T}_-, 
$$ 
while the other tau-function $\rho$ is given by 
$$ 
\rho_{\bf 0}=(y_+^2-y_-^2)\tau_{\bf 0}^2. 
$$ 
By applying the BTs  $R,S,T$ the general solution of the system 
is obtained at all the intersection points of the lines 
$\mathcal{L}_1,\mathcal{L}_2$ in terms of two solutions $y_\pm$ of P2.          
\subsection{Intersection of  $\mathcal{L}_1$ and  $\mathcal{L}_3$} 
 
A one-parameter family of mixed rational-Airy solutions may be generated 
at these points by applying the BTs to the seed solution 
$$ 
L_1=-2z, \quad L_2=0=X_1, \quad X_2=Y, \quad Y'+Y^2=2z 
$$ 
at the point ${\bf p}=(2,0)^T$. 
For example, applying the translational BT $T$ yields a corresponding 
one-parameter solution at the point ${\bf p+c}=(4,2)^T$, 
$$ 
L_1=\frac{4}{Y}-\frac{4z^2}{Y^2}, \quad 
L_2=2z-\frac{4z^2}{Y^2}, \quad 
X_1=Y-\frac{2z}{Y}, \quad 
X_2=\frac{1}{z}-{2z}{Y}. 
$$ 
The free parameter comes from the 
solution of the Riccati equation for $Y$, which  is
linearized to Airy's equation (with suitable scaling), viz 
$$ 
Y=(\log \phi)', \qquad \phi''-2z\phi=0. 
$$ 
The associated tau-functions are 
$$ 
\tau_{\bf p}= \exp (-z^3/6),  
\quad \rho_{\bf p}= \phi\,\exp (-z^3/3),   
\tau_{\bf p+c}=\phi ' \exp (-z^3/6),  
\quad \rho_{\bf p+c}= \phi z\exp (-z^3/3). 
$$ 
All solutions obtained from this by application of the BTs are 
rational functions of $z$ and $Y$. 

\subsection{Intersection of  $\mathcal{L}_2$ and  $\mathcal{L}_3$} 
 
Another one-parameter family of mixed rational-Airy solutions 
is found at the $\mathcal{L}_2,\mathcal{L}_3$ intersection points 
starting from the seed solution 
$$ 
L_1=L_2=-J^2-z, \qquad X_1=X_2=-J, \qquad J'+J^2=-z 
$$ 
at the point ${\bf q}=(1,1)^T$. So for instance, at 
${\bf r}=TRSR\cdot{\bf q}=(3,1)^T$ we find the solution 
$$ 
L_1=-z-\frac{2}{J}-\frac{z^2}{J^2}, \quad L_2=-z-\frac{z^2}{J^2}, \quad 
X_1=\frac{z}{J}, \quad X_2=\frac{z}{J}+\frac{1}{J^2+z}. 
$$ 
The Riccati equation for $J$ is linearized to 
a rescaled Airy's equation, 
$$ 
J=(\log \psi)', \qquad \psi''+z\psi=0. 
$$ 
The corresponding tau-functions at these points are 
$$ 
\tau_{\bf q}=\psi, \qquad 
\rho_{\bf q}=1, \qquad 
\tau_{\bf r}=\psi ', \qquad 
\rho_{\bf r}=\left|\begin{array}{cc} \psi & \psi ' \\ 
\psi ' & \psi '' \end{array}\right|. 
$$ 

\section{Acknowledgements} 

This poster was presented at the NEEDS conference at the Isaac Newton 
Institute, 25th-31st July 2001. I would like to thank the INI for 
supporting my stay in Cambridge during the Programme on Integrable Systems.

\end{document}